\documentstyle[twocolumn,aps,prb,psfig]{revtex}

\begin{document}

\title{Diagonal charge stripes giving a suppression of spectral weight 
at selected spots on the Fermi surface} 
\author{N.L. Saini$^1$, A. Bianconi$^1$, A. Lanzara$^1$, 
J. Avila$^2$, M.C. Asensio$^2$, S. Tajima$^3$, G.D. Gu$^3$, 
N. Koshizuka$^3$}

\address{$^{1}$
INFM Unit , Dipartimento di Fisica, Universit\`a "La Sapienza" -  
00185 Roma, Italy} 
\address{$^{2}$
LURE, Bat 209D Universite Paris-Sud, F-91405 Orsay, France} 
\address{$^{3}$
Superconductivity Research Laboratory,  ISTEC, Tokyo  135 , Japan.} 

\date{\today} 
\maketitle 
\medskip

In our Letter \cite{Saini} we have reported 
a direct evidence for charge stripes in 
diagonal direction in Bi2212. We have recorded the photointensity distribution 
in the $k$ space to probe the spectral weight 
within 50 $meV$ around the Fermi level. 
This Fermi surface image has been obtained by rotating the sample around its
 normal, while the detector is moved in previous $ARPES$ measurements 
for the similar information \cite{Ding}. 
Our experimental set-up allows us to probe directly the
 Fermi surface anisotropy while in the other approach the effects of intrinsic
 anisotropy are mixed-up with variation of matrix element effects. 
The colinearity of the sample normal and the rotation axis is within 
$\pm±0.5^{\circ }$, 
established by laser alignment and photoelectron diffraction of core levels.
 The Fermi surface image shows broken segment with hot spots 
(photointensity cusps of spectral weight) and missing parts 
(minimum of spectral weight) that indicate charge stripes in the $CuO_2$ plane. 
The clear experimental evidence for stripes is given by the large suppression
 (about $70\%$) of spectral weight at four points $P_n$ ($n$=1,2,3,4) 
on the Fermi surface in the first Brillouin zone. 
This suppression arises for coupling 
of electrons with a $1D$ charge density wave in the diagonal direction of 
wavevector $q\sim (0.4\pi ,0.4\pi )$ (see the white arrows in the 
upper panel of Fig. 1). 
This wavevector is two times the wavevector of $1D$ lattice fluctuations 
$Q\sim (0.2\pi ,0.2\pi )$ detected by anomalous x-ray diffraction and $EXAFS$
\cite{Bianconi}. 
Moreover the points $P_1,P_3 (P_2,P_4)$ are connected by the wavevector of the 
antiferromagnetic correlations $G(\pi ,\pi )$ in the stripe direction, 
(see the blue arrows in Fig. 1) indicating the additional coupling of 
electrons with spin density waves.

Mesot et al. \cite{Mesot} report 
a one-electron band calculation of the Fermi surface 
that is in qualitative and quantitative disagreemnt with the experimental data.
 In their calculations they include the contributions  of $umklapp$ satellite 
bands displaced by ±$\pm Q$ wavevector perpendicular to $\Gamma -X$ 
(due to diffraction from the superstructure in the $Bi-O$ plane) and satellite 
$shadow$ bands displaced by $G$ wavevector. This simple calculation reproduces 
some features of the experimental photointensity distribution, for example 
the suppression along the $\Gamma -Y$ 
direction due to matrix element effects, and 
the weak features due to $umklapp$ satellites. The calculated image shows hot 
spots where the main, $umklapp$ and $shadow$ band are crossing.
This calculation is unable to reproduce the missing parts on the main Fermi 
surface arising from the suppression of spectral weight 
(not to be confused with the temperature dependent opening of partial gaps). 
To make it more clear we show the photointensity higher than $50\%$ of the 
maximum, in the first Brillouin zone, in the lower panel, together with 
the calculated  curves for the main, $umklapp$ and $shadow$ bands. 
It is evident that the points $P_n$, where the large suppression of spectral 
weight occur, lie at the four crossing points of the main and the $shadow$ 
bands. Moreover these points can be located at the crossing point of the main 
Fermi surface and the lines connecting the $M-M_1$ points. In fact the charge 
density wavevector $q (0.4\pi , 0.4\pi )$ shown in Fig. 1 
gives the suppression of 
spectral weight along these lines. The calculation of Mesot et al. gives 
maxima at the points $P_n$ and is unable to account the main point of our work. 
To reproduce the observed experimental features one needs to make more 
sophisticated calculations as recently reported by Bansil et al 
\cite{Bansil} for angle 
scanning photoemission images including the superlattice of quantum stripes 
in the $CuO_2$ plane \cite{Bianconi}.
 Coming to the point of new electronic states reported in ref.\cite{Saini1}, 
the asymmetry of the $\Gamma -M$ and $\Gamma -M_1$ 
direction is an intrinsic effect 
and not due to sample misalignment. Unfortunately no systematic efforts 
were made to investigate the asymmetry along the two $Cu$-$O$-$Cu$
 bond directions 
prior to our measurements. In fact, our recent $Cu$ $K$-edge x-ray absorption 
measurements \cite{Saini2}
ensure that the asymmetry of the $\Gamma -M$ and $\Gamma -M_1$ directions 
is intrinsic of the $CuO_2$ plane. Moreover in Fig. 3 of ref.\cite{Saini1} 
it can be seen 
that the intensity of this band is about $33\%$ of the main and it is clearly 
stronger than the intensity of the umklapp satellite $(22\%)$. 
In addition the overlapping of the two photointensity polar cuts in the 
two directions in that figure rule out directly the argument of suspected 
sample misalignment.

FIGURE CAPTION

FIG.1 Plot of the experimental photointensity around the 
Fermi surface (upper) and the main Fermi surface in the first Brillouin zone 
after cutting off the low intensity features due to satellites (lower). 
The points $P_n$ of highest suppression of spectral weigth are indicated

\end{document}